\documentclass[sn-nature]{sn-jnl}

\usepackage{natbib}
\usepackage{graphicx}%
\usepackage{multirow}%
\usepackage{amsmath,amssymb,amsfonts}%
\usepackage{amsthm}%
\usepackage{mathrsfs}%
\usepackage[title]{appendix}%
\usepackage{xcolor}%
\usepackage{textcomp}%
\usepackage{manyfoot}%
\usepackage{booktabs}%
\usepackage{algorithm}%
\usepackage{algorithmicx}%
\usepackage{algpseudocode}%
\usepackage{listings}%
\usepackage{soul}
\usepackage{txfonts}
\usepackage{color}
\usepackage{siunitx}
\usepackage{caption}
\usepackage{hyperref}
\usepackage[hyphenbreaks]{breakurl}
\newcommand{\km}{{\rm ~km}}

\newcommand{\s}{{\rm ~s}}
\newcommand{\h}{{\rm ~h}}
\newcommand{\yr}{{\rm ~yr}}
\newcommand{\Myr}{{\rm ~Myr}}

\newcommand{\YORP}{{\rm YORP}}
\newcommand{\col}{{\rm col}}
\newcommand{\damp}{{\rm damp}}

\usepackage{soul}
\usepackage[textsize=tiny]{todonotes}

\begin{document}


\title[Confined tumbling state as the origin of the excess of slowly rotating asteroids]{Confined tumbling state as the origin of the excess of slowly rotating asteroids}

\author*[1]{\fnm{Wen-Han} \sur{Zhou}}\email{wenhan.zhou@oca.eu}

\author[1,2]{\fnm{Patrick} \sur{Michel}}

\author[1,3]{\fnm{Marco} \sur{Delbo}}

\author[4]{\fnm{Wenchao} \sur{Wang}}

\author[5]{\fnm{Bonny Y.} \sur{Wang}}

\author[6]{\fnm{Josef} \sur{\v{D}urech}}

\author[6]{\fnm{Josef} \sur{Hanu\v{s}}}

\affil[1]{\orgdiv{Laboratoire Lagrange, Observatoire de la C\^ote d'Azur}, \orgname{Universit\'e C\^ote d'Azur}, \orgaddress{ \city{Nice},  \country{France}}}

\affil[2]{\orgdiv{Department of Systems Innovation}, \orgname{University of Tokyo}, \orgaddress{ \city{Tokyo}, \postcode{10587}, \country{Japan}}}

\affil[3]{\orgdiv{School of Physics and Astronomy}, \orgname{University of Leicester}, \orgaddress{\city{Leicester}, \country{UK}}}

\affil[4]{\orgdiv{Department of Physics}, \orgname{University of Hong Kong}, \orgaddress{\city{Hong Kong}, \country{China}}}

\affil[5]{\orgdiv{Center for Computational Astrophysics}, \orgname{Flatiron Institute}, \orgaddress{\city{New York}, \country{USA}}}

\affil[6]{\orgdiv{Charles University, Faculty of Mathematics and Physics}, \orgname{Institute of Astronomy}, \orgaddress{\city{V Hole\v{s}ovi\v{c}k\'ach 2, 180\,00 Prague}, \country{Czech Republic}}}

\abstract{
The rotational distribution of asteroids as a function of their size is used {as a diagnostic of} their physical properties and evolution. Recent photometric surveys from the Gaia mission, allowing observation of asteroids with
long spin periods (for example $\geq 24$~h), found an excessive group of slow rotators and a gap separating them from faster rotators, which is unexplained by current theories. Here we developed an asteroid rotational evolution model capable of reproducing the observed distribution. {We suggest that this distribution is regulated by the competition between collisions and internal friction dampening of "tumblers" -asteroids with unstable rotation vectors, and that the slow rotator group is mainly populated by tumblers.} {We constrain the product of the rigidity and quality factor, which relates to the body's viscosity, to $\mu Q \sim 4 \times 10^9~$Pa. This number, two orders of magnitude smaller than the one assumed for monolithic boulders,} implies that {rubble pile} asteroids could have a porous structure or a thick regolith layer, and undergo stronger tidal effects.}

\maketitle

\section{Introduction}

It was initially assumed that asteroid rotation distribution had to take a Maxwellian form, peaking at about four revolutions per day as the result of the collision-induced random distribution of their spin vectors in three-dimensional velocity space \cite{Harris1979}. However, as observations progressed, an excess of fast rotators near the spin barrier (2.2~hours) \cite{Pravec2000, Pravec2008} and of slow rotators was noted for small asteroids \cite{Farinella1981, Dermott1984, Binzel1989, Fulchignoni1995, Pravec2000, Harris2002, Pravec2008, Pal2020}. The excess of fast rotators has been modeled \cite{Pravec2008} as due to spin up until the spin barrier by the Yarkovsky–O'Keefe–Radzievskii–Paddack (YORP) effect, which is a thermal torque influencing asteroids’ rotation rates over a long term \cite{Rubincam2000, Vokrouhlicky2003}. The YORP effect has indeed been observed for eleven asteroids \cite{Lowry2007, Taylor2007, Durech2022, Durech2024}.

On the other hand, the abnormal excess of slow rotators is not captured by the prevailing model involving the YORP effect and collisions \cite{Pravec2008}. Prior studies have modeled this excess empirically, either by artificially reducing the spin acceleration by a factor of 2 \cite{Pravec2008} or ceasing the rotation \cite{Bottke2015} beyond a certain period, or assigning a spin rate where lower rates are more probable \cite{Holsapple2022}. Yet, these methods do not rest upon a solid physical foundation. Moreover, recent observations \cite{Durech2023} reveal an obvious drop in the number density in specific size-dependent periods, forming a visible ``gap'' that separates the slow rotators from the faster rotators (Fig.~\ref{fig1}).

Another puzzle relates to the asteroids in non-principal rotation states, termed “tumblers” \cite{Harris1994}. Observation shows that nearly all observed tumblers are distributed in the slow rotation zone \cite{Harris1994, Pravec2005}. The distribution of these tumblers is constrained by a transition line fitting a power-law on period-diameter diagram \cite{Harris1994, Pravec2005, Pravec2014}, which coincidentally matches the newly discovered ``gap'' (Fig.~\ref{fig1}). A plausible explanation for the distribution of tumblers is still lacking \cite{Pravec2014}, especially for its correlation with the visible gap in the spin distribution of asteroids.

\section{Mechanism}

\subsection{Formation of the slow rotators and gap}
We constructed a self-consistent rotational evolution model that takes into account collisional excitation, internal friction damping, and the YORP effect on tumblers (see details in Methods). In this model, the tumbling motion can be initiated either by the YORP torque spinning down the asteroid to a quasi-static rotational state \cite{Vokrouhlicky2007}, or by sub-catastrophic collision, the latter occurring on a timescale:
\begin{equation}
\label{eq:tau_col}
    \tau_\col = \tau_{\rm col,0} \left( {D \over 1~\km} \right)^{(4\alpha - 10)/3} \left( {P \over 8\h} \right)^{(1-\alpha)/3}.
\end{equation}
Here $\tau_{\rm col,0} = 113$~Myrs is determined by the collisional frequency of main belt asteroids (Methods) and $\alpha = 3.2$ is the power index of the size distribution of main belt asteroids \cite{Holsapple2022}. In our simulation, we do not take into account the evolution of $\alpha$ {(otherwise $\alpha \leq 3.2$ \cite{Bottke2005} since most of asteroids were born big \cite{Morbidelli2009})}, {because} the spin distribution reaches an equilibrium state on a relatively short timescale ($\sim 300~$Myrs) {compared to the time scales of the evolution of the size-frequency distribution}. The timescale above is shorter when the asteroid rotates slower, indicating that tumbling is more easily triggered for slow rotators. The tumbling motion can be damped to the principal-axis rotation by internal energy dissipation on a timescale:
\begin{equation}
\label{eq:tau_damp}
    \tau_\damp = \tau_{\rm damp,0} \left( {D \over 1~\km} \right)^{-2} \left( {P \over 8\h} \right)^3,
\end{equation}
where $\tau_{\rm damp,0}$ depends on the {viscosity}
(Methods). Here, the period $P$ is the main lightcurve period, closely associated with the total angular momentum \cite{Pravec2014}, which is conserved during the damping process.

\begin{figure*}
    \centering
    \includegraphics[width = \textwidth]{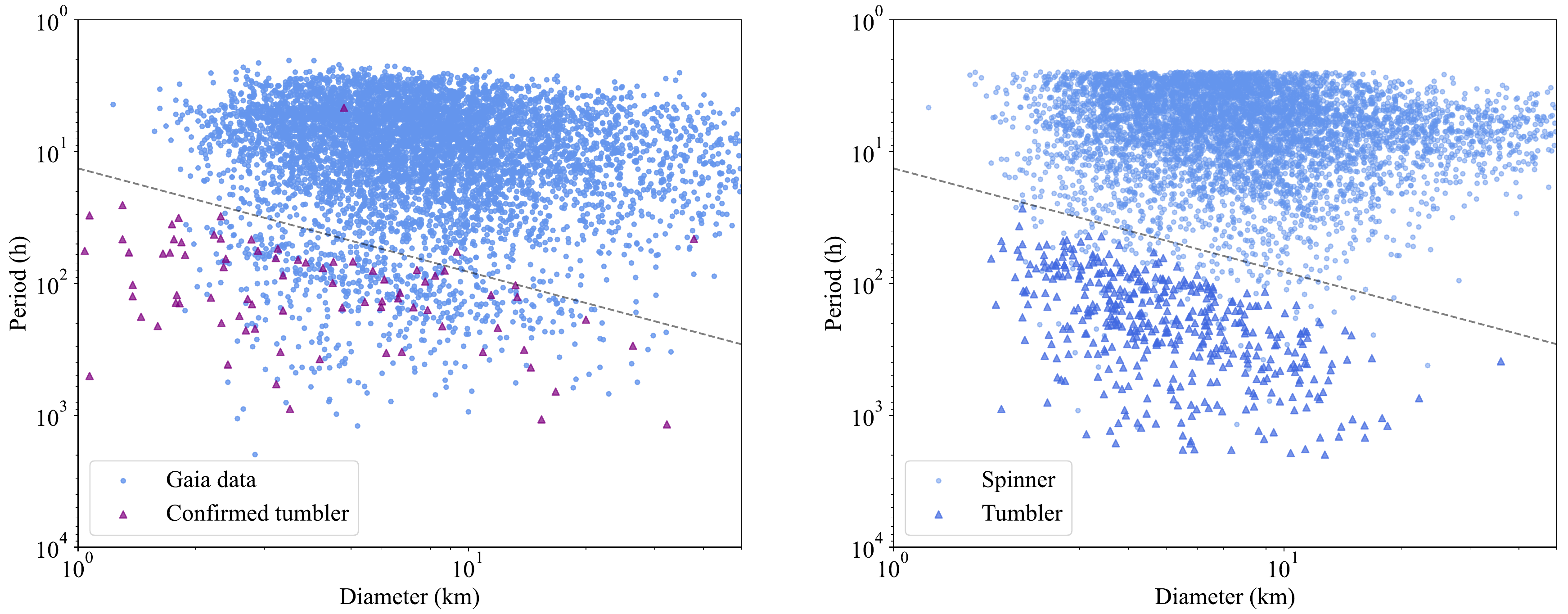}
    \caption{Period–diameter distribution: comparison of Gaia observations and simulations. Left: Observational data from Gaia \cite{Durech2023} showing the period–diameter distribution for asteroids, where the tumblers are identified using data from Asteroid Lightcurve Data Base (LCDB) \cite{Warner2009}. Right: Results from numerical simulations of the period–diameter distribution, performed using a strength parameter of $\mu Q = 4 \times 10^9~$Pa, a YORP weakening factor $f_{\rm weaken} = 0.1$, and assuming 50\% positive YORP torques. The grey line represents a fitted line that identifies the gap in the distribution (see equation \ref{eq:gap}).}
    \label{fig1}
\end{figure*}

These slow tumblers could evolve into a principal-axis rotation, a stable tumbling state with a fixed period, or a completely chaotic state (the nutation angle reaches $90^\circ$) \cite{Breiter2015}, depending on the initial orientation and the asteroid shape. In a completely chaotic state, the YORP effect could be ineffective, since the radiation torque can be averaged out over a long term if its direction is random. The same logic is used
to justify the absence of the binary YORP effect in non-synchronous asteroid binary system \cite{Cuk2005}. Therefore, we reasonably assume that a tumbler has a probability $p_{\rm fix}$ to maintain its period. In the long-term evolution, the YORP torque is reset numerous times after collisions due to the crater-induced YORP effect \cite{Bottke2015,Zhou2022, Zhou2024} {given the extreme sensitivity of YORP to small-scale topography \cite{Statler2009}. As a result, the tumblers experience random interchanges between the tumbling states both with or without a fixed period.} Therefore, the expected value of the YORP torque equals the standard YORP torque multiplied by a ``weaken factor'' $f_{\rm weaken} = 1-p_{\rm fix}$ which is smaller than 1 (Methods). The rotational evolution modes for asteroids in various rotational states are listed in Table 1.

\begin{figure*}
    \centering
    \includegraphics[width = \textwidth]{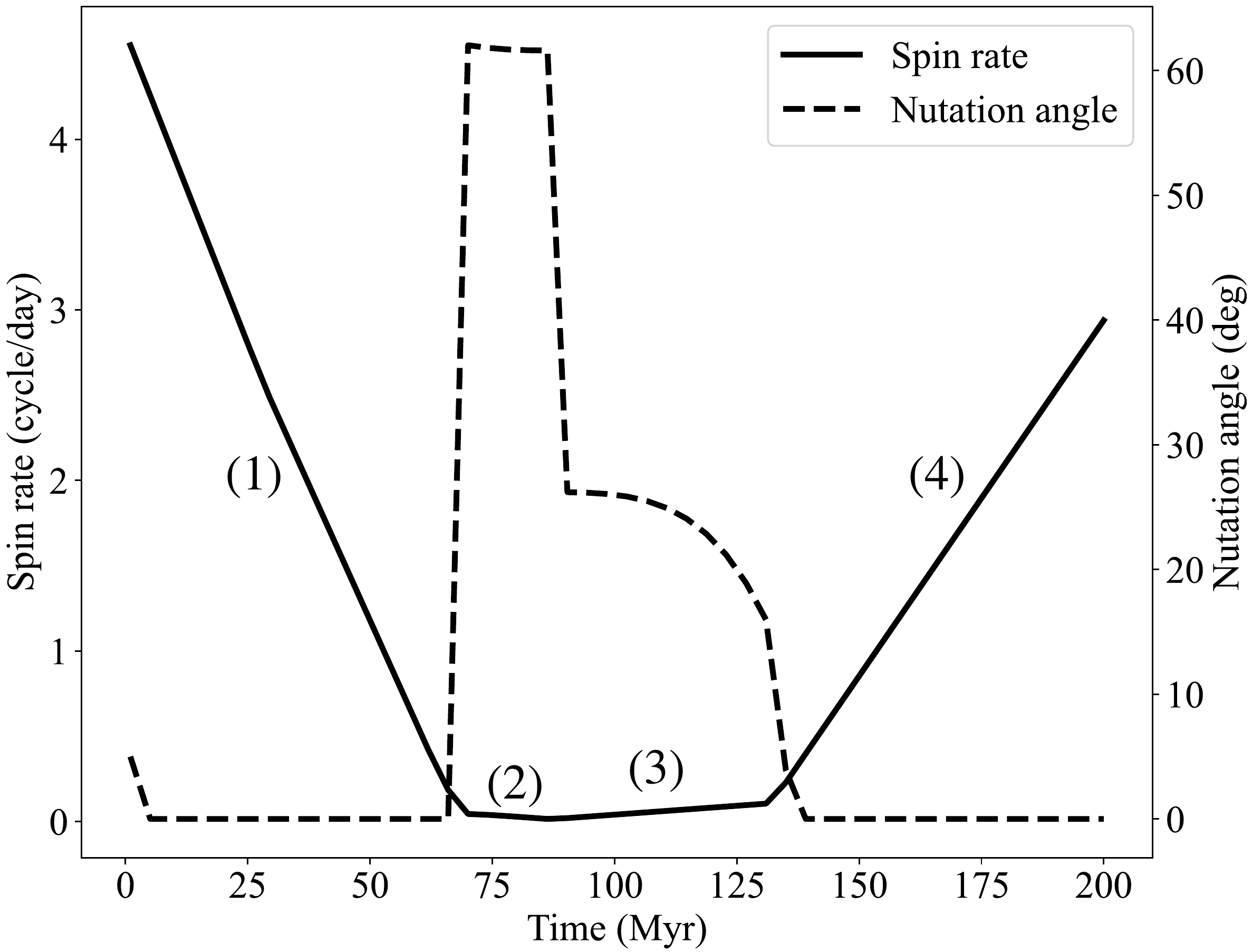}
    \caption{Rotational evolution of a synthetic asteroid {over 200 Myr}. The spin rate and nutation angle are denoted by the solid and dashed lines, respectively. This asteroid follows such a typical sequence: (1) it spins down, under the YORP effect, until it goes through a sub-catastrophic collision; (2) subsequently, the tumbling motion is triggered and it spins down at a slower rate than before due to a weak YORP effect until a new tumbling state is triggered; (3) then it starts to spin up at a slow rate until the tumbling is damped; (4) it spins up at a normal YORP acceleration until getting disrupted. It can be seen that the time fraction {of lifetime} in the slow region for asteroids is relatively high compared to that in the faster region, resulting in a larger number density of asteroid population in the slow region.}
    \label{fig2}
\end{figure*}

\begin{figure}
    \centering
    \includegraphics[width = \textwidth]{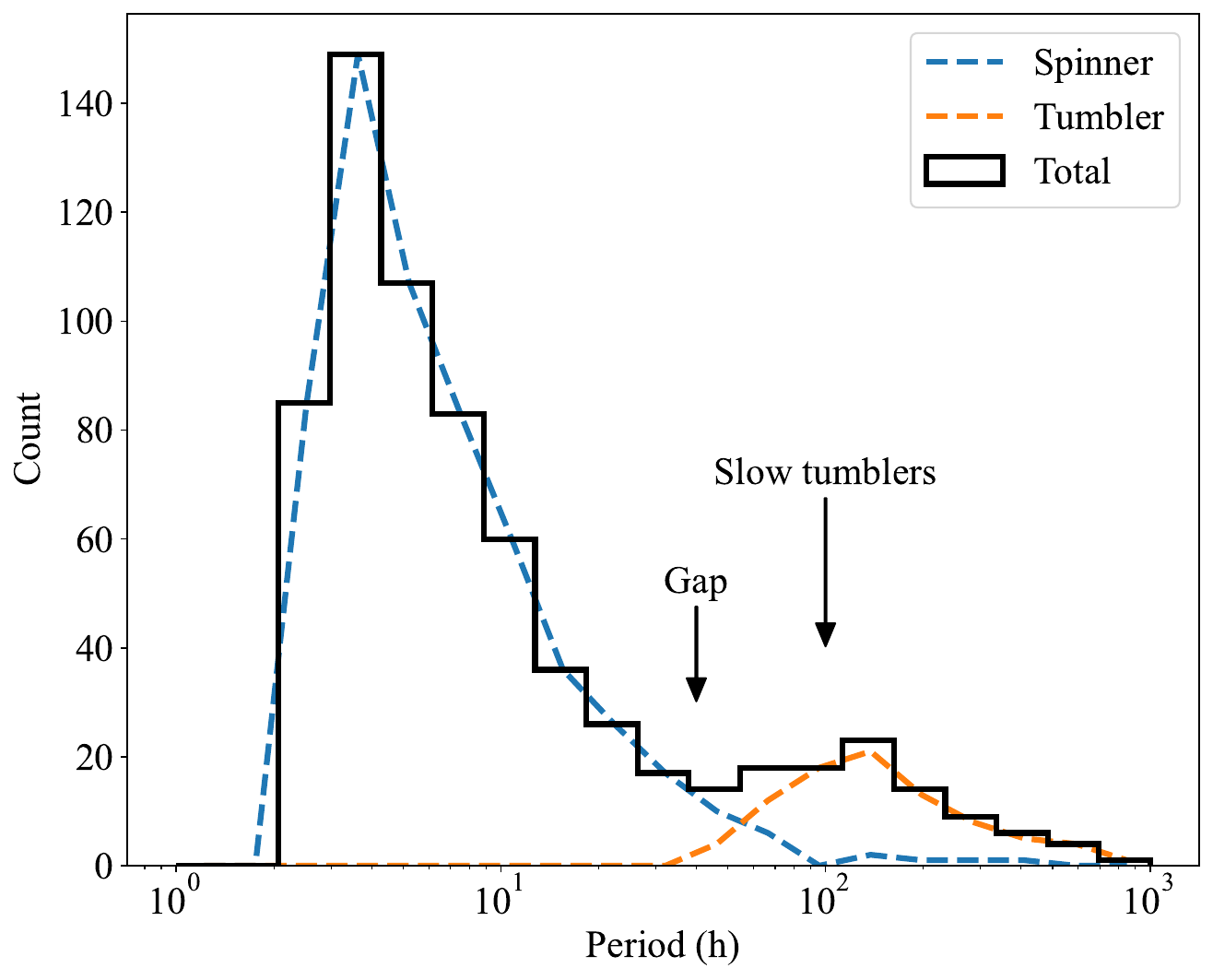}
    \caption{Bimodal period distribution for simulated asteroids between $3~\km$ and $4~\km$ as an example showing the location of the gap. The group of fast rotators is dominated by pure spinners while the group of slow rotators is dominated by tumblers. A distinct gap clearly separates the two groups. The peak of the distribution of slow tumblers is described by Eq.~\ref{eq:transition}. }
    \label{fig3}
\end{figure}

As a result of a weakened YORP torque, these tumblers evolve slowly in the long-period region (Fig.~\ref{fig2}), forming a concentrated group (Fig.~\ref{fig3}). The terminal spin distribution of our simulation is shown in Fig.~\ref{fig1}. The animation for the spin evolution can be found in Supplementary Video 1. Therefore, our results suggest that the slowly evolving tumblers are the origin of the excess of slow rotators. We compare our simulation result with the spin distribution for small asteroids between 3~km and 15~km reported by previous research \cite{Pravec2008}(Supplementary Fig.~\ref{fig8}) and constrain $f_{\rm weaken} \sim 0.1$, {assuming the YORP torque has equal probability of spinning up or down as indicated by the theoretical study \cite{Capek2004}.} Fitting the fraction of slow rotators in Gaia data gives $f_{\rm weaken} \sim 0.1$ (Methods). Our model predicts that most slow rotators are tumbling, consistent with observation data \cite{Pravec2005, Pravec2014}. 

\begin{figure*}
    \centering
    \includegraphics[width = \textwidth]{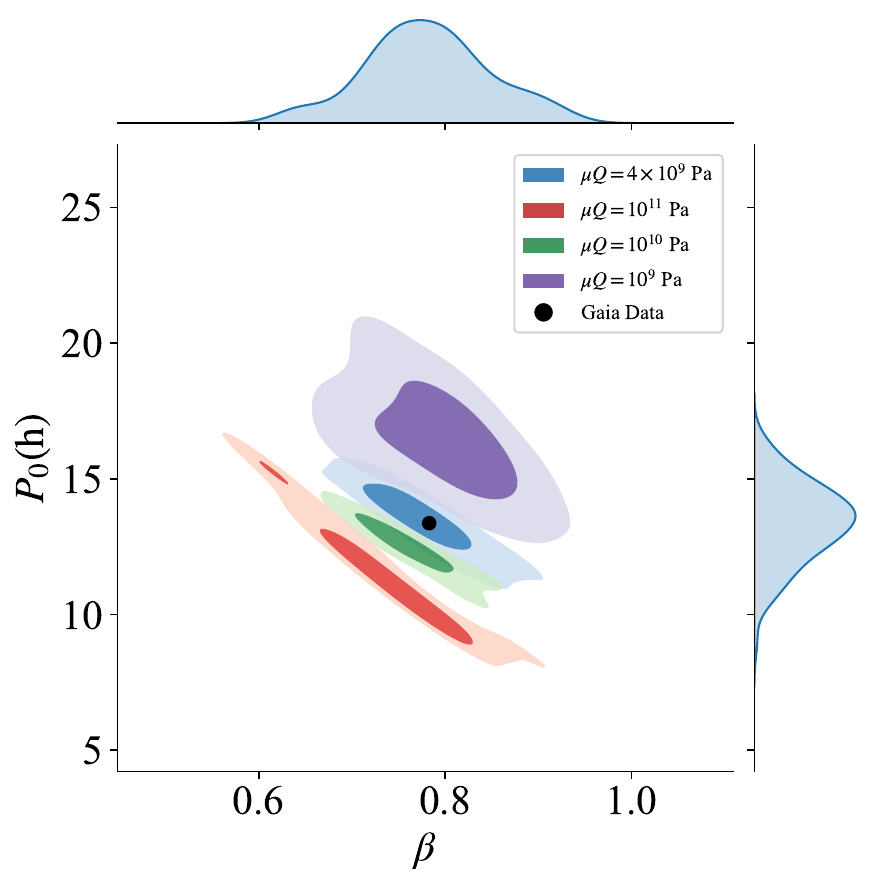}
    \caption{Probability density distribution of the gap parameters $P_0$ and $\beta$ (see Eq.~\ref{eq:gap}) for our simulation results with different $\mu Q$ (colored). The Gaia data is represented by the black dot. The parameters $\mu$ and $Q$ are obtained by the semi-supervised machine learning method (Methods).  We run 100 Monte Carlo simulations of rotational evolution for $\mu Q = 10^9,~4\times 10^{9},~10^{10}~$and $10^{11}~$Pa. The commonly assumed value $\mu Q$ for the damping of tumbling is $10^{11}~$Pa \citep{Pravec2014}. The two levels of the contours of the simulation results represent the 30\% and 68.27\% ($1\sigma$) probability of the data to lie below the contour. The curves (blue) on the marginal axes represent the density distribution of the parameter $\beta$ and $P_0$ in the case of $\mu Q = 4\times 10^{9}~$Pa.}
    \label{fig4}
\end{figure*}

\subsection{Location of the gap and its implications}

The distribution of tumblers is constrained by the transition line where the collisional excitation of tumbling balances the internal energy damping. The transition line can be obtained by equating Eq.~\ref{eq:tau_col} to Eq.~\ref{eq:tau_damp}:
\begin{equation}
\label{eq:transition}
    P = 8 \h \left( {\tau_{\rm col,0} \over \tau_{\rm damp,0}} \right)^{3/(8+\alpha)} \left( {D \over 1 \km} \right)^{\beta}  ,
\end{equation}
with $\beta =(4\alpha-4)/(8+\alpha) $. In the zone below this line, the excitation from collisions is dominant over damping, leading to the creation of numerous slowly rotating tumblers. However, we note that this transition line is not equivalent to the gap line but should be slightly lower than the gap line, as the tumblers
may undergo minor diffusion during damping. Considering $\alpha = 3.2$ \cite{Holsapple2022}, the slope of the transition line (Eq.~\ref{eq:transition}) becomes $\beta \simeq 0.785$. {The rarity of observed tumblers above the gap is consistent with our prediction, that is, the damping timescale above the gap is shorter than the excitation timescale. There is an outlier{, asteroid 1994 XF$_1$,} which is tumbling but located well above the gap \citep{Oey2017}, suggesting a recent collision event. Our calculations estimate that the most recent sub-catastrophic collision could have occurred less than four thousand years ago. {This timescale is comparable to that of space weathering \citep{Matsumoto2018}, highlighting this asteroid as a potential candidate for future spectral observations.}} More observations could be performed for asteroids with yet unmeasured rotation states in the long-period zone in order to further test our hypothesis that the large majority of them are tumblers. 

The observed gap in the period-diameter diagram (Fig.~\ref{fig1},\ref{fig3}), which separates the slow rotators from faster rotators, is the boundary of these slow tumblers. We used a semi-supervised machine-learning method to locate the gap of the observation data from Gaia {(see Methods section: Identification of the gap by a semi-supervised machine-learning method)}, which gives
\begin{equation}
\label{eq:gap}
    P = P_0 \left( {D \over 1\km} \right)^{\beta_0}.
\end{equation}
with $P_0 = 13.38 \h$ and $\beta_0 = 0.783$ for the observed gap. This is closely aligned with the predicted slope (i.e. 0.785) of the distribution of tumblers. 

In our model, the location of the gap and the transition line depends on the parameter $\tau_{\rm damp,0}$, {which is proportional to the ratio of the quality factor and Love number $Q/k_2$. In the literature, the Love number is commonly assumed to be inversely proportional to the asteroid rigidity $\mu$, although rigidity plays only a marginal role in dissipation, as compared to viscosity.} {A high $k_2$ means the body is easily deformed and} a low $Q$ value signifies that the body is relatively good at dissipating tidal energy and is not highly elastic \cite{Murray1999}. This provides invaluable information on the asteroid's internal structure and composition (e.g., the layered structure and the thickness of the regolith \cite{Nishiyama2021}). Furthermore, it is often associated with the tidal response of a celestial body, affecting the long-term evolution of the rotation and orbit of a binary system (e.g. binary asteroids \cite{Goldreich2009}, Earth-Moon system \cite{Murray1999}). However, due to the lack of seismic data or in-situ detection of tidal effects by space missions to asteroids, the value of $Q/k_2$ is poorly constrained despite{,} its importance in our understanding of the physical properties and evolution of asteroids. Our model offers a novel approach to constrain $Q/k_2$ by fitting the gap in the observed period-diameter distribution diagram of asteroids. We assume $\mu Q$ as a constant for our considered size range (1-50~km). {By fitting the gap in the simulation results to that in Gaia data {(Fig.~\ref{fig4})},} our best-fit model {suggests $\mu Q \sim 4 \times 10^{9}~$Pa, or equivalently $Q/k_2 \sim 5\times 10^8 (D/\rm km)^{-2}$, which produces a gap with the parameters $P_0 = 13 \pm 1~$h and $\beta_0 = 0.78 \pm 0.06$. This leads to $\tau_{\rm damp,0} \sim 0.4 \Myr$ in Eq.~\ref{eq:tau_damp}.} Our obtained value of $\mu Q$ is much smaller than usually assumed $ > 10^{11}~$Pa \cite{Pravec2014, Breiter2015, Fraser2018} for monolithic boulders {or $10^{13}~$Pa for cold less-porous solid minerals,} indicating that rubble piles are weaker (e.g. have a high porosity or a thick regolith layer \cite{Nishiyama2021}) and more susceptible to the tidal effect {than previously assumed. This} leads to a faster evolution \cite{Murray1999} and a larger equilibrium separation of binary asteroids \cite{Jacobson2011}.

\section{Methods}

\subsection{Identification of the gap by a semi-supervised machine-learning method}
{To identify the location of the gap from a statistical perspective, we use a semi-supervised machine-learning method to classify asteroid data into two groups and obtain the boundary line (the gap) that separates them.} {For a set of data,} we {performed pseudo labeling} by manually defining a grey zone {as} a starting reference zone for the model to locate the line, and label the data above the grey region as class I, and those below as class II, as shown in Supplementary Fig.~\ref{fig5}.
We use these well-classified data to fit a gap, using the linear Support Vector Machine (SVM) model \cite{Cortes1995}, which aims at finding the maximum distance between the gap and two groups by using stochastic descent to converge on the solution. {Expressed in the log scale, the gap is described by a line with the slope (k) and the intercept (b) with $k = \beta$ and $b = \log P_0$.} We train the model by using the following semi-supervised approach:
\begin{enumerate}
  \item Initialize the configuration of the gap line in log-scale: $k=0.6$, $b=1.2$. The y-distance of the gap is $0.4$. 
    The initial values are reasonably chosen by {observation}. The principle of choosing the values is to make sure the type of the asteroids above the gap is different from the type below the gap. {We leave the data points inside the grey area as unclassified data.}
  \item Use the manually classified data to train the linear SVC model and get the parameters of the model. 
    The parameters include $k$ and $b$.
  \item Use the new values of $k$ and $b$ obtained from the previous step to re-label the data points. 
    Similarly, the new $k$ and $b$ represent another gap and we label the data above the gap as class I, the data below the gap as class II {and leave data inside the gap as unclassified}.
  \item Redo step 2 by re-training the model until the gap gets converged, i.e., the values of $k$ and $b$ {have little variations} during each iteration.
\end{enumerate}

{We apply the above semi-supervised machine-learning method to both the observation and our simulation data. As the gap between slow rotators and fast rotators is clearly observed in Gaia data for the first time \cite{Durech2023}, we use the Gaia data reported by \citet{Durech2023} as the primary source of observation asteroids. All selected asteroids are main-belt asteroids, with semimajor axes ranging from 1.85 to 3.97 au.} The converged solution is $P_0 = 13.38~$hours and $\beta = 0.783$, as shown in Supplementary Fig.~\ref{fig5}. The obtained $\beta= 0.783$ for Gaia data closely matches the theoretical value $0.785$. {We run 100 Monte Carlo simulations on the rotational evolution model for each $\mu Q$, and then input the simulation results of the period-diameter distribution to the machine-learning model to identify the gap. The gap parameters obtained in simulations are shown in Fig.~\ref{fig4}.} However, we should note that due to the data quality {and the systematic uncertainty of the SVM model}, results might vary, resulting from specific parameters in the machine-learning model—such as the gap width or the initial value of $k$. We also use unsupervised clustering methods including but not limited to K-Means, Density-based spatial clustering of applications with noise (DBSCAN) and spectral clustering, as well as the geometric data analysis mentioned in \citet{Contardo2022}.  {In our experiments, unsupervised methods are more sensitive to their hyper-parameters and would introduce considerable uncertainty if they were forced to identify a single gap. An example of DBSCAN method with different min samples is shown in Supplementary Fig.~\ref{fig6}.} {In general, these unsupervised methods} do not yield {converging} results for our problem. {Animations of the iteration process with different initial conditions are presented in Supplementary Videos 2 and 3.}

\begin{figure*}
    \centering
    \includegraphics[width = \textwidth]{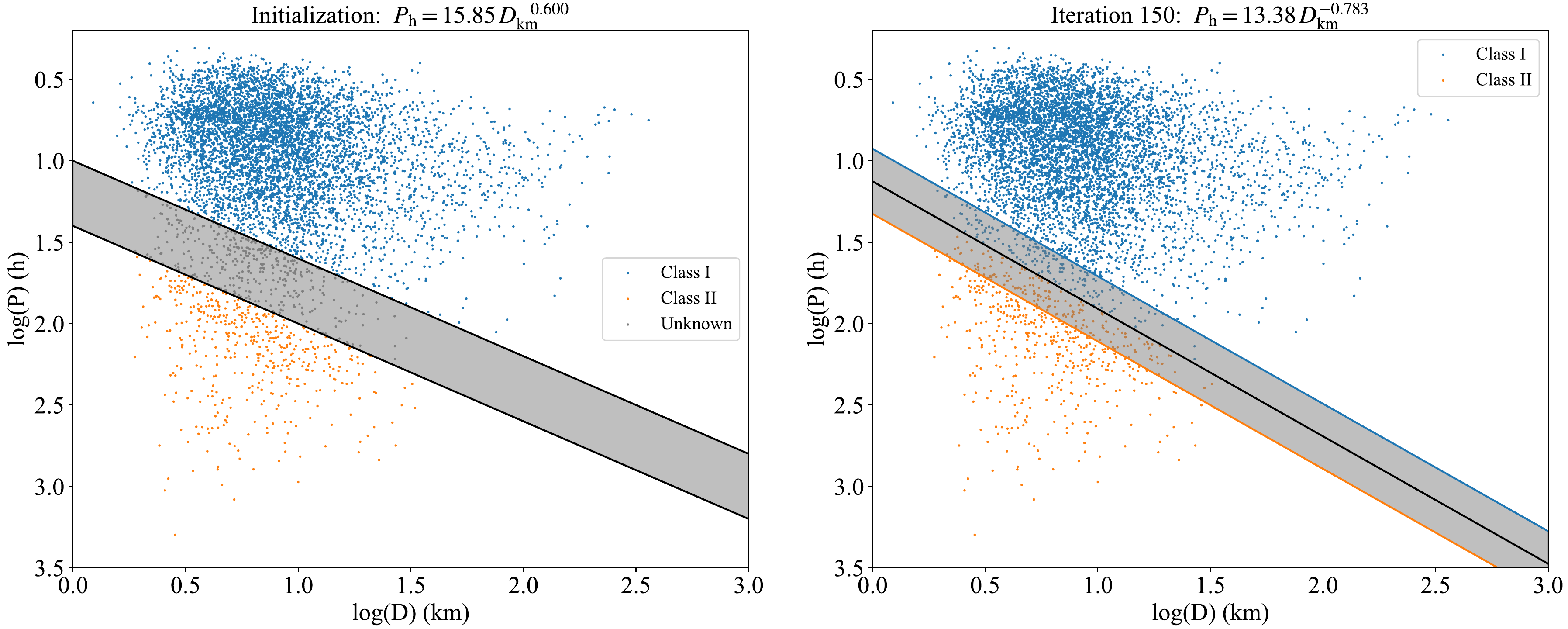}
    \caption{{Identification of the gap. \textbf{a}. The gap is initialized manually with a shaded area. The dots above and below the shaded area are classified as fast rotators (class I) and slow rotators (class II), respectively, while the dots in the shaded area are unclassified. The y-distance of the shaded area is 0.4. \textbf{b}. The converged result for the gap is shown after 150 iterations (Methods).}}
          \label{fig5}
\end{figure*}

\begin{figure*}
    \centering
    \includegraphics[width = \textwidth]{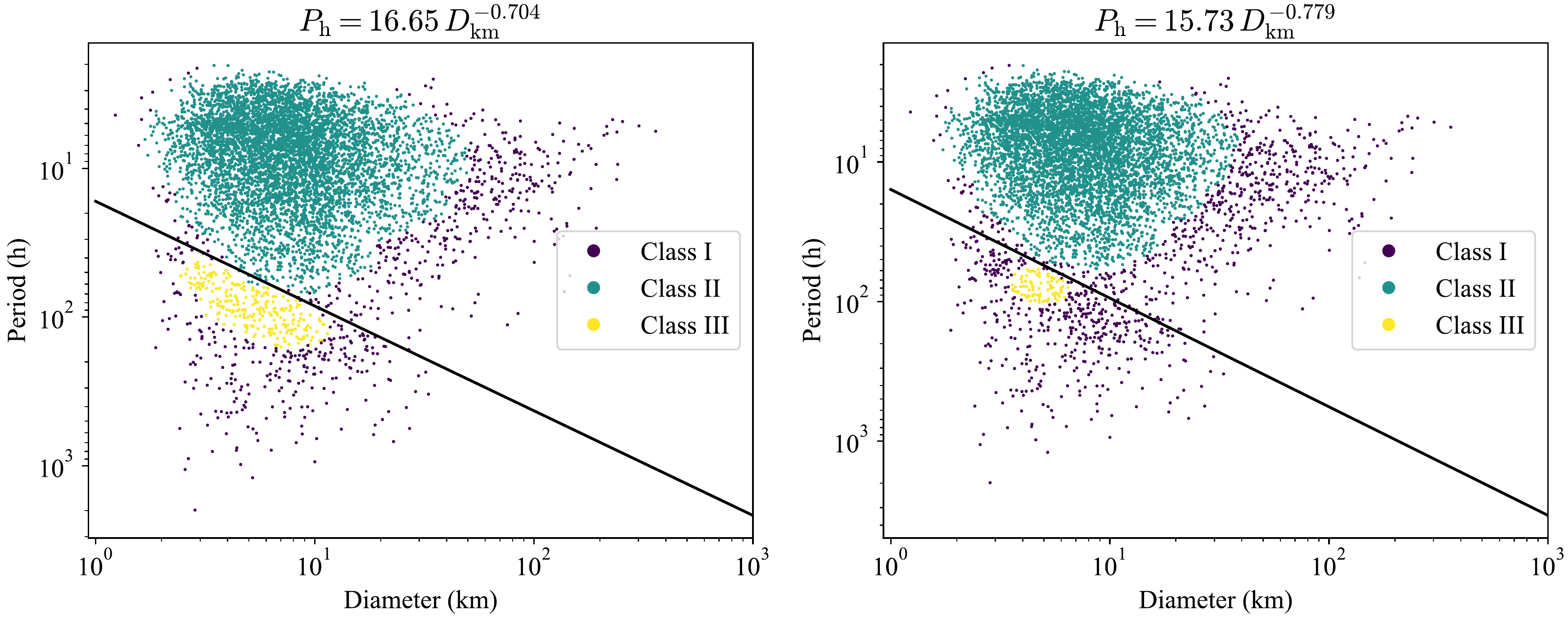}
    \caption{Classification by the DBSCAN method with min sample number of \textbf{a}, 40 and \textbf{b}, 50.}
    \label{fig6}
\end{figure*}

    

    

\subsection{Rotational evolution model}

The long-term rotational dynamics of asteroids are mainly dominated by the YORP effect, which is a radiative torque caused by radiation from the irregular asteroid surface, and sub-catastrophic collisions.
Both collisions and the YORP effect can transfer positive or negative angular momentum, leading to spin-up or spin-down of asteroids. Comparing the collision timescale with the YORP timescale, it is obvious that the YORP effect dominates over collisions unless the asteroid enters an extremely slow rotation. The term "YORP cycle" is used to describe the process by which the asteroid evolves into an end-state where its rotation state needs resetting under the YORP effect \cite{Bottke2015}. One YORP cycle is terminated when (1) the spin rate exceeds the rotational disruption limit $\omega = \sqrt{4 \pi G \rho / 3} \sim 2.2~\rm hours$; (2) a sub-catastrophic collision occurs; and (3) the spin rate decreases to zero. 

These three cases yield distinct outcomes (Table \ref{tab1}). In the first case, the asteroid is assumed to be reshaped or divided into fragments. The spin rate should be reset, as well as the YORP torque since it is sensitive to the surface topology \cite{Statler2009, Zhou2022}.
We consider the new object to maintain a pure spin state, as the rotational disruption chiefly alters the angular momentum along the major principal axis. Moreover, the process of reshaping or uniting the rubble piles rapidly dissipates energy to realign the asteroid. Thus, the nutation angle, which gauges the degree of tumbling, remains nearly zero.
In the second case, a sub-catastrophic collision induces the spin rate to renew with a random increment and the obliquity to reset randomly. Since a collision lacks a preferred direction of incidence, it imparts an angular momentum with equal probability to all directions, thereby instigating tumbling. 
In the third case, the asteroid spins down to a quasi-static rotational state, triggering a tumbling state due to the YORP effect or collisions. In our model, the initial spin rate distribution follows a Maxwell distribution with the peak at 8 hours, the minimum at 2000~hours, and the maximum at 2.2 hours. In fact, the initial condition does not affect significantly the spin distribution at equilibrium. We generate {as many test particles as asteroids in the Gaia sample with the same sizes and orbital elements} and run the simulation for 400~Myrs. The spin distribution arrives at equilibrium after around 300~Myrs.

\subsection{Sub-catastrophic collisions}
Sub-catastrophic collisions transfer angular momentum to asteroids, modifying the spin rate and potentially triggering tumbling.
We follow the route of \citet{Farinella1998} to calculate the characteristic collision timescale at which an asteroid's rotational state is reset. To transfer the angular momentum that is comparable to that of the target asteroid with the angular speed $\omega_{\rm targ}$ and radius $r_{\rm targ}$, the radius of the impactor is
\begin{equation}
\label{eq:r_imp}
    r_{\rm imp} \sim \left( { \omega_{\rm targ} r_{\rm targ} \over  v_{\rm imp} }    \right)^{1/3} r_{\rm targ},
\end{equation}
where $v_{\rm imp} \sim 5.8 \km / \s$ is the average impact speed between asteroids \cite{Bottke1994}. The collision timescale for such a collision can be estimated as
\begin{equation}
\label{eq:tau_col2}
    \tau_{\rm col} = { 1  \over P_{\rm i} N_{r>r_{\rm imp}} (r_{\rm imp} + r_{\rm targ})^2 }.
\end{equation}
Here $P_{\rm i} \sim 2.85 \times 10^{-18} \km^{-2} \yr^{-1} $ is the intrinsic collisional probability \cite{Bottke1994}, and $N_{r>r_{\rm imp}}$ is the number of the asteroid that has a larger radius than $r_{\rm imp}$:
\begin{equation}
\label{eq:N_imp}
    N_{r>r_{\rm imp}} = N_0 \left( {r_{\rm imp} \over r_0} \right)^{1-\alpha}
\end{equation}
with $-\alpha$ being the power index of the size-frequency distribution (SFD) of asteroids. Substituting Eq.~(\ref{eq:r_imp}) and Eq.~(\ref{eq:N_imp}) into Eq.~(\ref{eq:tau_col2}), we can obtain 
\begin{equation}
\begin{aligned}
    \tau_{\rm col} & \simeq { f_{\col}\over P_{\rm i} N_0} \left( {1\over v_{\rm imp} r_0^3} \right)^{(\alpha - 1)/3} \omega_{\rm targ}^{(\alpha-1)/3} r_{\rm targ}^{(4\alpha - 10)/3} \\
    &.  
\end{aligned}
\end{equation}
{where $f_\col $ is a dimensionless coefficient related to the shape of asteroids, the impact angle and the density ratio of the impactor to the target.} The general form of the collision timescale we use in our simulation is 
\begin{equation}
    \tau_\col = \tau_{\rm col,0} \left( {D \over 1~\km} \right)^{(4\alpha - 10)/3} \left( {P \over 8\h} \right)^{(1-\alpha)/3} . 
\end{equation}
{Applying $f_\col \simeq 0.83$, $\alpha = 3.5$, $r_0 = 1~$km and $N_0 = 3.5 \times 10^5$ for a rough estimate, \citet{Farinella1998} obtain $\tau_{\rm col,0} = 50.8~$Myrs. With an update to $f_\col$ by \citet{Farinella1999}, the hereafter studies utilise $\tau_{\rm col,0} = 228.8~$Myrs.
\citet{Bottke2005} show the wavy characteristics of the size distribution of asteroids, in particular, the decreasing slope in the small size end. To account the slightly shallower distribution in the size range (0.1, 1)~km, \citet{Holsapple2022} uses $N_0 = 6.1 \times 10^5$ and $\alpha = 3.2$ and obtains $\tau_{\rm col,0} = 113$~Myrs. In this study, we follow the size distribution fitted by \citet{Holsapple2022}.}
In our numerical code, when the evolution time reaches $\tau_\col$, the spin rate $\omega$ receives an increment $\Delta \omega$ due to a random collision:
\begin{equation}
    \Delta \omega = (\sqrt{2+2\cos \gamma}  - 1)\omega
\end{equation}
where $\gamma$ is the random angle between the projectile angular momentum and the target angular momentum, uniformly distributed from 0$^\circ$ to $360^\circ$. After such a sub-catastrophic collision, the nutation angle is set to be a {random number between $0^\circ$ and $90^\circ$}, denoting a tumbling state. We also reset the YORP coefficient, since the YORP torque is shown to be sensitive to the change of fine surface structures (e.g. boulders and craters) \cite{Statler2009, Golubov2012, Zhou2022}.

\subsection{YORP effect}
The YORP effect is a radiative torque produced by the recoil force of the emitted photons from the asteroid. The asteroid could spin up or spin down under the YORP effect. In our model, the asteroid is initially assigned with a YORP torque, which can be expressed as \citep{Nesvorny2007, Mysen2008, Golubov2019, Marzari2020}
\begin{align}
    T_{{\rm YORP}, z} & = \delta_\YORP f_\YORP {\Phi R^3 \over c} (\cos 2\epsilon + {1 \over 3}) \\
    T_{{\rm YORP}, \epsilon} & =  - f_\YORP {2\Phi R^3 \over 3c} \sin 2\epsilon.
\end{align}
{Here $T_{{\rm YORP}, z}$ and $T_{{\rm YORP}, \epsilon}$ are the YORP torque components that modify the spin rate and the obliquity, respectively. The obliquity $\epsilon$ ranges from $0^\circ$ to $90^\circ$.}  Here $\Phi$ is the mean flux of solar radiation at the asteroid's orbit, $R$ is the asteroid radius and $c$ is the light speed.

{The dimensionless YORP coefficient $f_\YORP$ is highly sensitive to the shape of asteroids. The probability distribution of $f_\YORP$ is commonly modeled as either a normal distribution \citep{Rossi2009,Marzari2011,Jacobson2014} or an exponential law \citep{Golubov2019}. In our study, we adopt a half-normal distribution to ensure $f_\YORP$ remains positive, setting the standard deviation at 0.005, which is a typical value \citep{Marzari2020, Zhou2022}. This simple YORP model leads asymptotically asteroid obliquity to either $0^\circ$ or $180^\circ$, consistent with observations on asteroid obliquities \citep{Vokrouhlicky2003}, although simulations show the asymptotic obliquity could be also $90^\circ$ for some asteroids \citep{Vokrouhlicky2002}. We introduce the parameter $\delta_\YORP$, which can take values of 1 or -1, to control the sign of $T_{{\rm YORP},z}$. A positive $\delta_\YORP$ signifies a spin-up torque, while a negative value indicates a spin-down torque. For asteroids with an unrealistic zero thermal inertia, simulations suggest a tendency to spin down \citep{Vokrouhlicky2002, Zhou2020}. Conversely, for realistic asteroids with non-zero thermal inertia, the YORP effect appears to equally drive spin-up or spin-down \citep{Capek2004}. To date, eleven confirmed YORP detections all show increasing spin rates \citep{Durech2024}, a finding that contradicts theoretical predictions and remains unexplained. Possible observational biases include: (1) all detections are for near-Earth objects (NEOs), primarily retrograde rotators from the $\nu_6$ resonance in the main belt; (2) all detections are for fast rotators, which are more likely to be spinning up towards to the spin barrier. In the literature, $\delta_\YORP$ is typically assumed to be equally likely to be positive or negative \citep{Rossi2009, Marzari2011, Hanus2011, Jacobson2014, Marzari2020}. We maintain this assumption in our primary analysis but also explore scenarios with a preference for either positive or negative $\delta_\YORP$. Two extreme cases, which may not be realistic but are helpful for a quick grasp of the idea, are shown in Supplementary Fig.~\ref{fig7} where $\delta_\YORP$ has 90\% probability to be 1 or -1. Generally, an increased likelihood of positive YORP torque leads to a reduction in the population of slow rotators, aligning with expectations. Observations, however, poorly constrain the proportion of slow rotators, as one can see from inconsistencies between data derived predominantly from an asteroid family \citep{Pravec2008} and data from Gaia observation on main belt asteroids \citep{Durech2023}. Consequently, this study does not endeavor to refine the parameters that define the fraction of slow rotators but rather focuses on examining the gap feature and the distribution of tumbling asteroids. Fortunately, the preference to spin up or down does not affect the location of the gap and the distribution of tumbling asteroids.}

\begin{figure*}
    \centering
    \includegraphics[width = \textwidth]{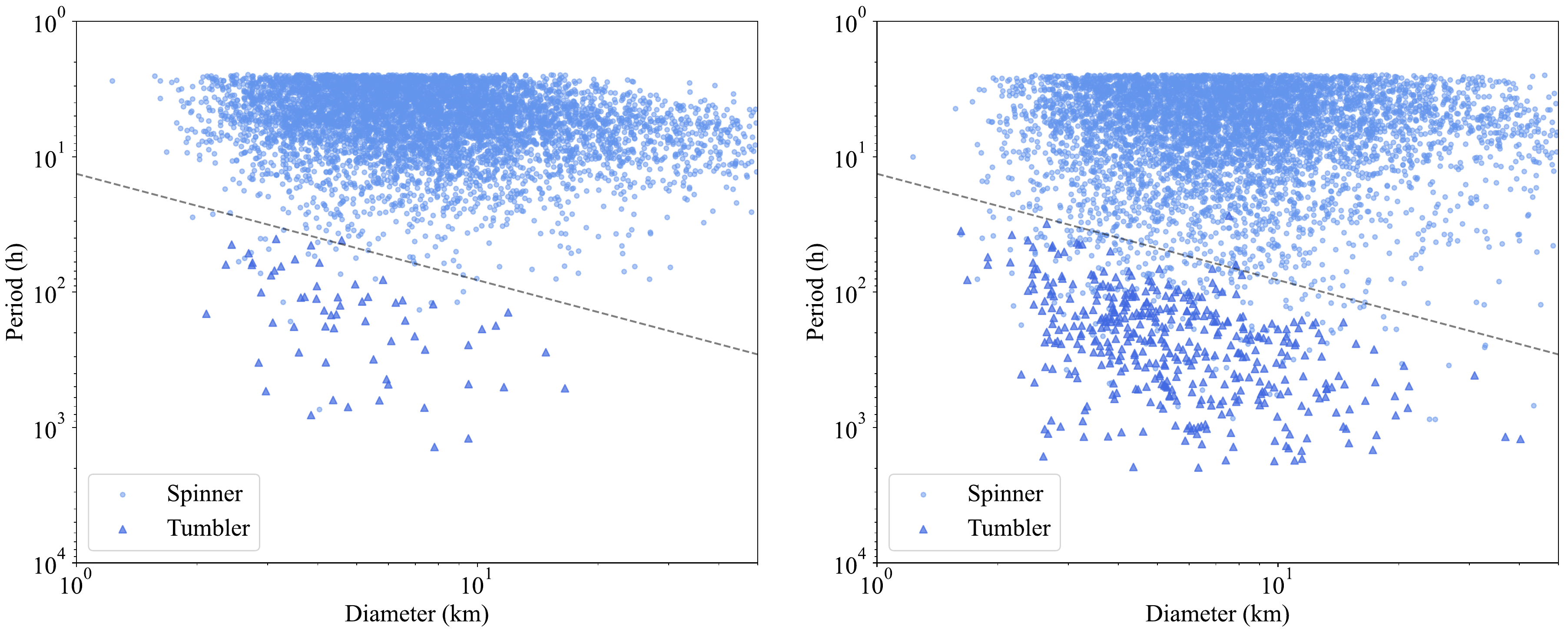}
    \caption{Results for \textbf{a}, 90\% positive YORP torques and \textbf{b}, 90\% negative YORP torques.}
    \label{fig7}
\end{figure*}

The YORP coefficient is reset when the asteroid goes through a rotational disruption (i.e. the spin rate exceeds the limit corresponding to a period of 2.2 hours), or experiences a sub-catastrophic collision, after which the surface is largely deformed, leading to a change on the YORP torque \cite{Statler2009}. If the asteroid spins down to a quasi-static rotation, the tumbling motion could be easily triggered by the YORP torque or collisions. In our model, we assign a nutation angle of $45^\circ$ to the asteroid if the period becomes longer than 1000 hours to denote the tumbling motion. The understanding of the YORP effect on a tumbling asteroid is still largely limited due to the shape-sensitivity of the YORP torque and the sometimes chaotic nature of tumbling motion \cite{Vokrouhlicky2007, Cicalo2010, Breiter2011, Vokrouhlicky2015}. 
\citet{Breiter2015} demonstrate that the combined impact of the YORP effect and the damping on tumblers might produce a stable tumbling motion with a fixed period, or can evolve into a completely chaotic tumbling motion when the nutation angle arrives $90^\circ$ \cite{Breiter2015}. In the latter case, the YORP effect might be ineffective since the radiation torque can be averaged out over a long term if the direction of the radiative torque is random. The same logic is used to justify the absence of the binary YORP effect in non-synchronous asteroid binary systems \cite{Cuk2005}. Therefore, a tumbling asteroid has a probability $p_{\rm fix}$ of maintaining its angular momentum under the complex coupling effects of the YORP torque and internal energy dissipation. The high sensitivity of $p_{\rm fix}$ to the initial rotation state and the shape of asteroids, makes it hard to obtain a typical value of $p_{\rm fix}$ theoretically, especially considering that the above studies have ignored the thermal inertia. 

In our model, the probability $p_{\rm fix}$ can be translated into a "weaken" factor of the YORP effect $f_{\rm weaken}$ in the simulation of long-term rotational evolution of asteroids by
\begin{equation}
    f_{\rm weaken} = 1 - p_{\rm fix},
\end{equation}
so that the expected value of the YORP torque is 
\begin{equation}
   {T_{\YORP}}' = f_{\rm weaken} T_{\YORP}.
\end{equation}
The rationale behind this is explained as follows: since the YORP torque is sensitive to the surface deformation \cite{Statler2009}, the YORP torque can be reset from time to time \cite{Bottke2015} by the crater-induced YORP (CYORP) torque\cite{Zhou2022, Zhou2024}. With every reset of the YORP torque, the tumbler has a probability of $p_{\rm fix}$ to fix the period (invalidate the YORP torque), the expected value of the YORP torque over a long term can be simply estimated as $(1-p_{\rm fix})T_{\YORP}$, given that the simulation time $t$ ($\sim$ hundreds of Myrs) is much larger than the reset timescale $\tau_{\rm reset}$ ($ \leq 1~$Myr) \cite{Bottke2015, Zhou2022}. In fact, the YORP effect that accounts for the random reset by the CYORP torque is referred to as the ``stochastic YORP'' effect \cite{Bottke2015}, which reduces the efficiency of the YORP torque \cite{Bottke2015, Zhou2022}. However, this effect makes no difference in the equilibrium spin distribution, because it works for all asteroids, indicating that it only affects the evolution rate. For the sake of simplicity, we choose to overlook this effect, especially considering the absence of a comprehensive model for the stochastic YORP effect \cite{Zhou2022}.

It is possible to constrain the value of $f_{\rm weaken}$ by observations, taking advantage of the fact that $f_{\rm weaken}$ directly influences the number of slow rotators. A greater $f_{\rm weaken}$ introduces a larger fraction of slow rotators. In case the Gaia observation has a bias against fast asteroids, we utilize the histogram provided by \citet{Pravec2008} (see their Fig.~2) of the spin distribution of 268 small main-belt and Hungaria asteroids with the diameter range 3–15 km, which was considered as the best estimate of the rotation rate distribution of small asteroids \cite{Bottke2015}. {$\sim 30\%$ of the asteroids in \citet{Pravec2008} belong to two asteroid families, i.e., the Hungaria and Phocaea families. Family member asteroids are thought to share the same age. Thus they are more suitable with agreeing with our synthetic asteroids that evolve concurrently in our simulations, compared to a general sample e.g. from Gaia data, the latter potentially suffering from a bias against the detection of low-lightcurve-amplitude fast rotators
.} We vary the value of $f_{\rm weaken}$ in our simulation and compare the resulting spin distribution of asteroids in the same size range with that in \cite{Pravec2008}. The most consistent result is shown in Supplementary Fig.~\ref{fig8}, indicating $f_{\rm weaken} \sim 0.1$. Compared to ground-observation, Gaia is less biased for slow rotators, while fast rotators might be more biased. By fitting the fraction of slow rotators ($\omega < 1~$cycle/day), we fit $f_{\rm weaken} \sim 0.12$. In the result presented in Fig.~\ref{fig1}, we adopted $f_{\rm weaken} = 0.1$. 

{We note that the unknown preference of the YORP torque to be positive or negative also affects the histogram, as shown in Supplementary Fig.~\ref{fig7}. The parameter $f_{\rm weaken}$ may be smaller than 0.1 if the YORP torque favors positivity. Consequently, the constraints on $f_{\rm weaken}$ become complicated due to inconsistent information from different observations and unknown chaotic behaviors of the YORP torque. However, fortunately, the value of} $f_{\rm weaken}$, {given $f_{\rm weaken} < 1$}, does not change the position and the slope of the gap in the slow-rotating region and the distribution of tumblers. {A more precise constraint on the parameter $f_{\rm weaken}$ may be deferred for future investigations, provided that unbiased observations demonstrate agreement with the histogram in the spin rate distribution.}

\begin{figure*}
    \centering
    \includegraphics[width = \textwidth]{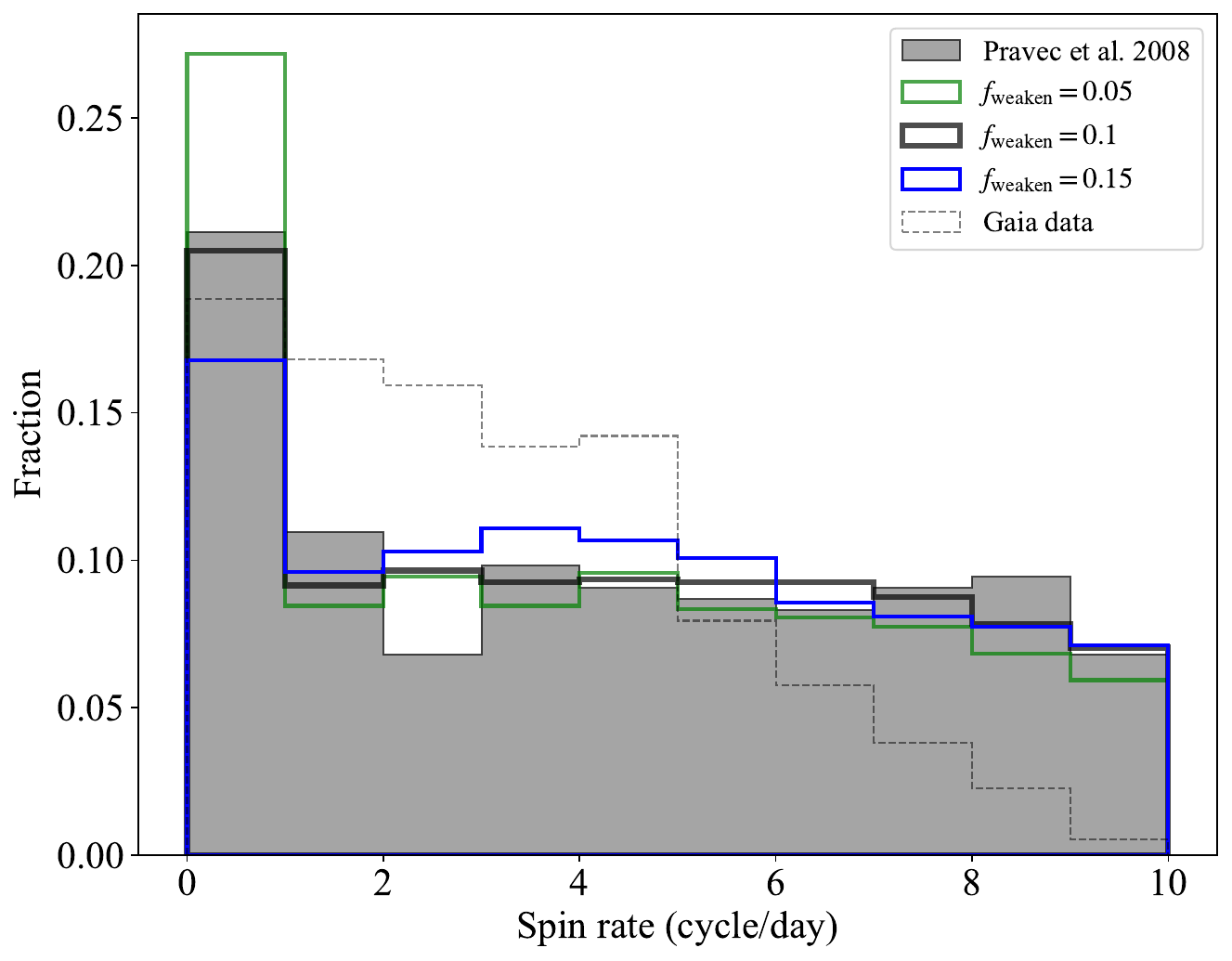}
    \caption{Comparison of the spin rate distribution of asteroids (3-15~km) from \citet{Pravec2008} and \citet{Durech2023} with our model results assuming $f_{\rm weaken}$ = 0.05, 0.1 and 0.2. As $f_{\rm weaken}$ decreases, the fraction of the slow rotators increases. Our findings indicate that $f_{\rm weak} = 0.1$ provides the best match to observation data by \citet{Pravec2008}, especially for the slow rotators with $\omega < 1~$cycle/day. The less consistency in the group of faster asteroids with Gaia data could arise from systematic issues in the Gaia dataset. Compared to ground-based observations, Gaia might exhibit a greater bias against fast rotators (the expected excess of fast rotators isn't evident), while the slow rotators are less biased. In fact, focusing only on the fraction of slow rotators ($\omega < 1~$cycle/day), we fit $f_{\rm weaken} \sim 0.12$ for Gaia data, which is close to the result fit from \citet{Pravec2008}.}
    \label{fig8}
\end{figure*}




\subsection{Internal energy dissipation}
Due to the internal friction, a tumbling object will finally evolve into a pure spin state around the major principal axis. The damping rate of asteroids is usually estimated by using an empirical quality factor $Q$ whose inverse measures the energy lost per precession period, {and the Love number $k_2$ describes the tumbling-caused deformation. In the classical theory \cite{Burns1973}, $k_2$ is translated into the rigidity $\mu$ for a homogeneous elastic body: }
\begin{equation}
    k_2 \simeq \frac{\pi G \rho^2 D^2}{19 \mu},
\end{equation}
{leading to the commonly used ``$\mu$Q`` prescription.}
The damping timescale is generally expressed as
\begin{equation}
\label{eq:tau_damp2}
    \tau_\damp \simeq {\mu Q \over \rho R^2 \omega^3} A = \tau_{\rm damp,0} \left( {D \over 1~\km} \right)^{-2} \left( {P \over 8\h} \right)^3,
\end{equation}
where $A$ is a dimensionless shape factor. The uncertainty of this estimate comes from the quality factor $Q$, which is usually set to be 100 but never actually measured for rubble pile asteroids \cite{Burns1973, Pravec2014}. The value of the shape factor $A$ is suggested to be 18 in Breiter's model \cite{Breiter2012, Pravec2014}, although its value varies by two orders of magnitude in the older literature \cite{Burns1973,Efroimsky2000,Molina2003, Sharma2005, Breiter2012, Caudal2023}. The nominal timescale is
\begin{equation}
    \tau_{\rm damp,0} \simeq 1.1~{\rm Myrs} ~{\left(\mu Q \over 10^{10}{\rm Pa} \right)}~{\left(2 {\rm g~cm^{-3}} \over \rho\right)}
\end{equation}

{While rubble piles have shear rigidity \cite{Goldreich2009}, dissipation and the damping rate are not very sensitive to its value. Recent studies have demonstrated that it is the viscosity that predominantly defines the efficiency of friction \cite{Efroimsky2015, Frouard2018, Kwiecinski2020}. An accurate approach to
the problem should be based on a rheological equation including the viscosity as a key
parameter. This treatment would render a frequency dependence of the quality factor.}

However, the viscosity of asteroids is still unknown and more complex rheologies may be needed \cite{Frouard2018,Kwiecinski2020}. {For convenient comparison with previous studies, we utilize the classic $Q/k_2$ method with the $\mu Q$ prescription in this work.} We assume $\mu Q$ as a constant for the size range of our interest (1-50~km), although $\mu Q$ could vary with the size \cite{Goldreich2009} or the frequency \cite{Nimmo2019}. In our model, the nutation angle $\theta$ evolves as 
\begin{equation}
    \dot \theta = 1/\tau_\damp
\end{equation}
for the sake of simplicity, although $\dot \theta$ should be a complex function of $\theta$ \cite{Sharma2005, Breiter2012}. Each time the asteroid is excited to tumble under the YORP effect or a collision, the nutation angle is reset to {a random number between 0$^\circ$ and $180^\circ$ for simplicity, considering the probability distribution of the nutation angle reset by a sub-catastrophic collision is unknown.}

{Our best fit model suggests that $\mu Q \sim 4 \times 10^9~$Pa (see Fig.~\ref{fig4}), which is equivalent to $Q/k_2 \sim 5 \times 10^8(D/{\rm km})^{-2}$. The results with $\mu Q \sim 10^{11}$~Pa and $10^9~$Pa are shown in Supplementary Fig.~\ref{fig9}. It can be seen that the usually assumed $\mu Q \sim 10^{11}~$Pa yields a gap above the observed gap, as a result of a weak damping effect. Conversely, a $\mu Q \sim 10^9~$Pa creates a gap lower than the observed one, according to diagnosis by our semi-supervised machine learning method.}

\begin{figure*}
    \centering
    \includegraphics[width = \textwidth]{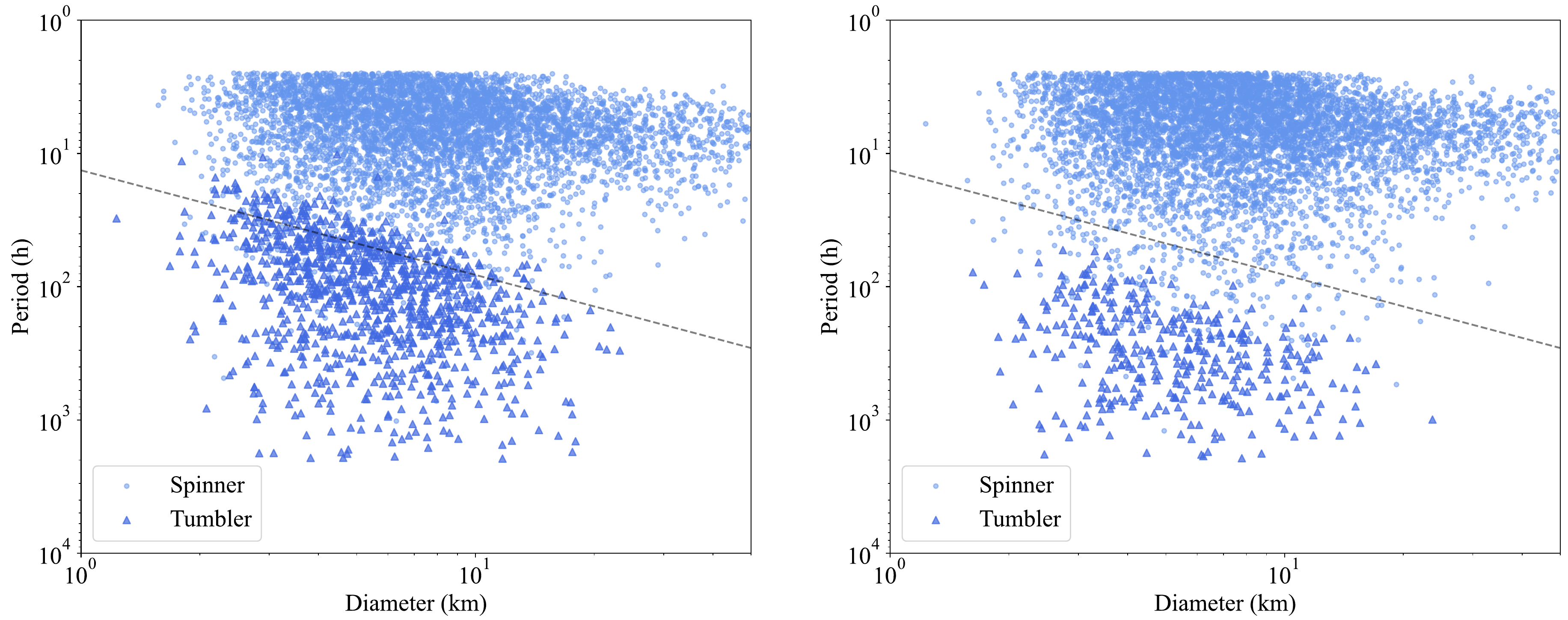}
    \caption{Period-diameter distribution of the simulation results with \textbf{a}, $\mu Q \sim 10^{11}$~Pa and \textbf{b}, $10^9~$Pa. The dashed line is the gap identified for Gaia data.}
    \label{fig9}
\end{figure*}

\bmhead{Data availability}
For the observation data, the code is available at https://github.com/WH-Zhou/Zhou-2024-Gap-Finder.

\bmhead{Code availability}
For the gap identification, the code is available at https://github.com/WH-Zhou/Zhou-2024-Gap-Finder.

\bmhead{Acknowledgments}
We are grateful for the constructive suggestions from reviewers. W.H. Zhou would like to acknowledge the funding support from the Chinese Scholarship Council (No.\ 202110320014). P. Michel acknowledges funding support from the French space agency CNES and from the European Union's Horizon 2020 research. The work of J. \v{D}urech and J. Hanu\v{s} was supported by the grant 23-04946S of the Czech Science Foundation.

\bmhead{Author contributions}
W.H.Zhou led the project. W.H.Zhou proposed the model, derived the formula, carried out the numerical simulations, analyzed the results, and led the writing of the paper. M.Delbo and W.H.Zhou initiated the collaboration.
W.C.Wang, Y.Wang, and W.H.Zhou led the gap identification using machine learning. M.Delbo and P.Michel contributed to writing the manuscript. J.\v{D}urech and J.Hanu\v{s} provided the asteroid data based on Gaia observations. All authors collaborated on the interpretation of the results. 

\bmhead{Competing interest}
Authors declare that they have no competing interests.

\bmhead{Tables}

\begin{table}[h]
\caption{Outcomes of YORP-cycle endstates.}\label{tab1}%
\begin{tabular}{@{}lllll@{}}
\toprule
Events & Spin rate  & Obliquity & YORP torque & Rotation mode \\
\midrule
Rotational disruption \footnotemark[1]    & Reset   & Remain & Reset  & Pure spin  \\
Quasi-static rotation   & Remain   & Remain & Reverse and weakened  & Tumbling  \\
Sub-catastrophic collision    & Reset   & Reset & Reset and weakened & Tumbling \\
\botrule
\end{tabular}
\footnotetext[1]{This applies to the larger remnant; disruption can also mean mass shedding.}
\end{table}


\end{document}